\input iopppt
\pptstyle
\title{On the Symmetries of the Edgar-Ludwig Metric}
\author{Alan Barnes}

\address{School of Engineering and Applied Science,
Aston University, Aston Triangle, Birmingham B4 7ET, UK
\footnote\dag{E-Mail: \ \ {\tt barnesa{\bf @}aston.ac.uk}
\ \ \ Fax:\ \  {\rm +44 121 333 6215}}}

\pacs{04.20}
\jl{6}
\submitted

\date

\beginabstract
The conformal Killing equations for the most general (non-plane wave) 
conformally flat pure radiation field are solved to find the conformal
Killing vectors.  As expected fifteen independent conformal Killing vectors
exist, but in general the metric admits no Killing or homothetic vectors.
However for certain special cases a one-dimensional group of
homotheties or motions may exist and in one very special case,
overlooked by previous 
investigators, a two-dimensional homethety group exists. No higher dimensional
groups of motions or homotheties are admitted by these metrics.
\endabstract
\vfill\eject

\section{Introduction}
The most general conformally flat pure radiation (or null fluid) field 
which is not a plane wave is given by (Ludwig and Edgar, 2000)
$$ ds^2 = (x V(x, y, u) - w^2) du^2 + 2x du dw - 2w du dx - dx^2 - dy^2 
\eqno(1)$$
with
$$V = N(u)(x^2 + y^2) +2 M(u)x + 2 F(u)y +2 S(u) \eqno(2)$$
where $M$, $N \ne 0$, $F$ and $S$ are arbitrary functions of the
coordinate $u$.  This metric generalises a solution found by Wils (1989).
Only three of these functions are essential as the form of the metric is   
preserved by the coordinate transformations
$$ d u = d\tilde u/\alpha(\tilde u)\qquad 
w = \alpha \tilde w  +\dot \alpha x \qquad
\tilde x = x \qquad \tilde y = y \eqno(3)$$
where $\alpha$ is an arbitrary non-zero function of $\tilde u$ and a
dot signifies a partial derivative with respect to $\tilde u$. Under
this coordinate transformation $V$ transforms as
$${\tilde V}(\tilde x , \tilde y , \tilde u ) = \alpha^{-2}(V(x, y, u) 
+2 x(\alpha \ddot \alpha - \dot \alpha ^2)) \eqno(4)$$

In the original form of the metric (Edgar and Ludwig, 1997), this
transformation was (in effect)  used to set $M = 0$.  However it is
more convenient to use the coordinate freedom to set $N =1$ as in
Edgar and Vickers (1999).
Thus, dropping tildes, the most general conformally flat pure
radiation field is given by the metric (1) with
$$V = x^2 + y^2 + 2 M(u)x + 2 F(u)y +2 S(u) \eqno(5)$$
where $M$, $F$ and $S$ are arbitrary functions of the coordinate $u$. 

It is interesting to study the symmetries of the Edgar-Ludwig metric
for a number of reasons.  Firstly the metric is of interest in its own
right as it has a physically realistic matter content, namely pure
radiation.  Being conformally flat the metric admits a fifteen
parameter group of conformal symmetries and it might be thought that,
for special values of the functions $M$, $F$ and $S$, some of these
symmetries would reduce to homotheties or isometries as happens in the
case of the conformally flat perfect fluid solutions found by Stephani
(1967). Although the most general Stephani metrics admit no
isometries, there is a rich set of special cases which admit
isometries; see  Barnes and Rowlingson (1990) and  Seixas (1992) for a
treatment of the non-expanding case and Barnes (1998) for the
expanding case.  By way of contrast in the conformally flat
null fluid case, the special cases admitting homotheties and
isometries are much more restricted except for the case of plane
waves. In this paper only the non-plane wave case will be considered
as the symmetry structure of plane waves solutions is well-known
(Ehlers and Kundt, 1962).   

Secondly metrics such as the Stephani and Edgar-Ludwig solutions
present a stiff test for implementations of the Karlhede algorithm
(Karlhede, 1980; Karlhede and MacCallum, 1982) as their classification
requires the calculation of the third and fourth derivatives of 
the curvature tensor respectively (Bradley, 1986; Koutras A, 1992;
Skea, 1997).  
In a recent paper Pollney et al.~(2000) presented a classification
of the Edgar-Ludwig metric using the GRTensor implementation of the
Karlhede algorithm.  However their results are not consistent with
those obtained by Skea (1997) using the implementation of the same
algorithm in CLASSI ({\AA}man, 1987). GRTensor predicts a
non-trivial isotropy group whereas CLASSI predicts no isometries
at all.  Moreover there appears to be a
bug in the GRTensor output routines as evidenced by the
form given for the real part of $(D^2 \Phi)_{4,4}$ in Pollney et
al.~(2000). Given this uncertainty it is useful to investigate the
symmetry structure by an alternative method, namely the direct
integration of conformal Killing's equations.  
Moreover this method unlike the standard version of the Karlhede
algorithm, gives expressions for the  conformal
Killing vectors directly. The results on isometries obtained in this paper 
are consistent with those obtained by the CLASSI version of the
Karlhede algorithm.  

Thirdly Ludwig and Edgar (2000) have developed a new formalism for the
investigation of the existence of homotheties and isometries. They
have used this formalism to investigate the symmetries of their 
metric and they plan to extend the method to handle
conformal Killing vectors.  Their formalism works well when only a
single homothetic or Killing vector exists. However, when several such
vectors exist, its use is somewhat controversial as evidenced by the
discussion following Edgar's talk on the method at the recent GR16
meeting in Durban.  Thus  it is useful to have available results
on the symmetry structure of this metric obtained by more tried and
tested methods. 

\section{Conformal Killing Vectors}
The conformal Killing equations are $\xi_{i;j} + \xi_{j;i} =2 \sigma g_{ij}$
or equivalently
$$g_{ij,k}\xi^k + g_{ik} \xi^k_{,j} + g_{kj}\xi^k_{,i} =2\sigma g_{ij}
\eqno(6)$$
A straightforward but somewhat lengthy calculation reveals that the
components of the conformal Killing vector $\xi^i$  and the associated 
conformal factor must have the form
$$\eqalignno {
\xi^u &= (a (x^2+y^2) + b y +d x + c)/x &(7)\cr
\xi^w &= 2a w^2 +w(2\beta y +\gamma +2a_u x - d_u) + H(u,x,y) &(8)\cr
\xi^x &= (a(x^2-y^2)-b y - c)w/x + 2\beta x y + \gamma x 
+a_u(x^2-y^2) -b_u y -c_u &(9)\cr
\xi^y &= (2a y + b) w + \beta (y^2 - x^2) +\gamma y + \epsilon 
+2a_u x y +b_u x &(10)\cr
\sigma &= 2a w +2a_u x +2\beta y + \gamma &(11)}$$
where $a$, $b$, $c$, $d$, $\beta$, $\gamma$ and $\epsilon$ are functions 
of the coordinate $u$ and where $u$ subscripts denote partial derivatives.

Equations (7)-(11) could be used to investigate the conformal symmetries
of any metric of the form (1) as their validity does not depend on the
precise form of the function $V$ in the metric.  The $xu$, $yu$
and $uu$ components of the conformal Killing equations do depend on the
precise form of $V$ and remain to be satisfied.  For the Edgar-Ludwig
metric where $V$ is given by equation (5), integration of these three
equations for $H$ gives 
$$\fl
H(x, y, w) = (a_{uu}-2a M)(x^2+y^2) +(b_{uu}-2b M)y +c_{uu}-2c M \cr
\lo- \frac{1}{2}a x^3 - a x y^2 -(a F + b/2)x y +k x \eqno(12)\cr
\lo- x^{-1} (\frac{1}{2}a y^4 + (a F + b/2)y^3 +( b F + a S + c/2)y^2
 +(c F + b S)y + c S)$$
where $k$ is a function of $u$ only. The functions $\beta$, $\gamma$,
$\epsilon$,  $a$, $b$, $c$, $d$ and $k$ must also satisfy the
following linear system of differential  equations: 
$$\eqalignno{
2\beta_u &= 2a F - b &(13)\cr
\gamma_u &= 2a S -c &(14)\cr
\epsilon_u &= b S -c F &(15)\cr
2a_{uuu} &= 2 a M_u +2\beta F +4 a_u M -2 d_u - \gamma &(16)\cr
b_{uuu} &= b M_u +2\beta S -d F_u +2b_u M -2d_u F -\epsilon &(17)\cr
c_{uuu} &= c M_u -d S_u +2M c_u -2 S d_u -F\epsilon + S\gamma &(18)\cr
2d_{uu} &= -6a S +2b F - 3c -2k &(19)\cr
2k_u &= -2a S_u -2d M_u +2a_u S -2b_u F +c_u -4 d_u M &(20)}$$
Using standard techniques this linear system may be reduced to a first
order linear system for fifteen unknowns, namely $a$, $b$, $c$ and
their first and second derivatives,  $d$ and its first derivative
together with $\beta$, $\gamma$, $\epsilon$ and $k$.  The general
solution of this system thus involves fifteen arbitrary constants of
integration and so there is a fifteen-parameter family of conformal
Killing vectors. This is to be expected as the metric
is conformally flat and so admits a conformal symmetry group of
maximal dimension. The algebra involved in deriving equations
(12)-(20) is rather heavy and has been 
checked\footnote\ddag{For full details see the files\ 
{\tt http://www.aston.ac.uk/{\~ \null}barnesa/el.red} and
{\tt el.log}.}
using the computer algebra system Reduce (Hearn, 1995). 

\section{Homotheties and Isometries}
We now consider whether the  metric can admit any
homothetic motions or isometries.  For a proper homothetic motion, the
conformal factor $\sigma$ appearing in equation (6) must be a non-zero
constant and for an isometry we have $\sigma = 0$.  
Hence from equation (11) it follows immediately that $a = \beta = 0$
and $\gamma = \sigma$. Since $\sigma$ is constant, equations
(13)-(15) imply that $b = c = \epsilon_u = 0$. Furthermore from (16)
and (17) it follows that $d = d_0 -\frac{1}{2} \sigma u$ and $k = 0$
where $d_0$ is a constant. From (12) it now follows that $H(x, y, u) = 0$.   

Equations (17), (18) and (20) then restrict the form of the metric
functions $F(u)$, $S(u)$ and $M(u)$ as follows
$$\eqalignno{
(2d_0 -\sigma u)M_u &= 2 \sigma M &(21)\cr
(2d_0 -\sigma u)F_u &= 2 \sigma F - 2 \epsilon &(22)\cr
(2d_0 -\sigma u)S_u &= 4\sigma S - 2 \epsilon F &(23)}$$
where $d_0$, $\epsilon$ and $\sigma$ are all constants.  
In general these equations will not be satisfied and so the
generic metric will not admit any homotheties or
isometries.  However, in special cases $F$, $S$ and $M$ will satisfy
these equations and the metric will admit one (or more) homotheties or
isometries.  

For an isometry $\sigma = 0$ and (since $\xi^i \ne 0$) it
follows that $d_0 \ne 0$ and so without loss of generality we may
rescale $\xi^i$ so that $d_0 = 1$. The solution of equations (21)-(23)
is
$$M(u) = m_0 \qquad F(u) = f_0 - \epsilon u \qquad 
S(u) = s_0 +\frac{1}{2}\epsilon^2 u^2 - f_0\epsilon u \eqno(24)$$
where $m_0$, $f_0$ and $s_0$ are arbitrary constants.
Without loss of generality we may set $f_0=0$ by means of the
coordinate transformation $\tilde u = u  -f_0/\epsilon$ if $\epsilon \ne 0$
or $\tilde y = y + f_0$ if $\epsilon =0$.
Thus the metric admits a Killing vector 
if and only if the function $V$ in the metric (1) can be
written as
$$V = x^2 + y^2 +2 m_0 x -2 \epsilon u y +2s_0 + \epsilon^2 u^2 \eqno(25)$$
The Killing vector has the form
$$\xi = \partial_u +\epsilon \partial_y \eqno(26)$$
This result agrees with that obtained by Ludwig and Edgar (2000) 
and is consistent with that of Skea (1997) who worked in a coordinate 
system in which $M=0$ rather than $N=1$ in equation (2).

For an homothety we may scale $\xi^i$ so that $\sigma = 1$ and then,
by the coordinate transformation $\tilde u = u - 2d_0$, we may set
$d_0=0$.  The solution of equations (21)-(23) in this case is
$$M = m_1 u^{-2}\qquad F = f_1 u^{-2} + \epsilon \qquad
S = s_1 u^{-4} +f_1\epsilon u^{-2} +\epsilon^2/2 \eqno(27)$$
where $f_1$, $m_1$ and $s_1$ are arbitrary constants.
Without loss of generality we may set $\epsilon =0$ by means
of the coordinate transformation $\tilde y = y + \epsilon$.
Thus the metric admits a homothetic vector 
if and only if the metric function $V$ in the metric (1) can be
written as
$$V = x^2 + y^2 +2 m_1 u^{-2} x +2f_1 u^{-2} y +2s_1 u^{-4}\eqno(28)$$
The homothetic vector has the form
$$\xi = -\frac{1}{2}u\partial_u +\frac{3}{2}w\partial_w
+x\partial_x +y\partial_y \eqno(29)$$
This result is also consistent with that obtained by Ludwig and
Edgar (2000) although they used a different coordinate system
from the one they used to investigate Killing vectors. In this
coordinate system the function $N$ appearing in $V$ in equation (2)
is not normalised to unity.
The use of different coordinate systems for the investigation of 
homotheties and isometries makes it difficult to determine whether
the spacetime can admit two (or more) Killing and homothetic vectors.
However the analysis is straightforward if the same coordinate system
is used  throughout;  equations (25) and (28) are satisfied
simultaneously if and only if
$$ V = x^2 + y^2 \eqno(30)$$
and in this case the metric (1) admits a Killing vector $\xi_0$ and a
homothetic vector $\xi_1$ given by
$$\xi_0 = \partial_u \qquad
\xi_1 = -\frac{1}{2}u\partial_u +\frac{3}{2}w\partial_w
+x\partial_x +y\partial_y \eqno(31)$$
In this case the (maximal) homothety group is two-dimensional.
The group is non-Abelian as the commutator of its generators is
$$[\xi_0\ \xi_1] = -\frac{1}{2} \xi_0 \eqno(32)$$
This case was overlooked by Ludwig and Edgar (2000) who concluded that
maximal dimension of the homothety group admitted by their metric was
one.

\section{Summary}
The conformal symmetries of the Edgar-Ludwig metric have been
investigated and an explicit form for the most general conformal 
Killing vector obtained. This vector depends on fifteen functions
of the coordinate $u$ which satisfy a first order linear 
differential system.  The most general Edgar-Ludwig metrics
which admit a Killing vector or a homothetic vector have been obtained; 
they depend on three arbitrary constants ($m_0$, $s_0$ and $\epsilon$ 
or $m_1$, $s_1$ and $f_1$ respectively) whereas the general metric
depends on three arbitrary functions of $u$.  Finally it has been shown
that there is a single metric, overlooked by previous investigators,
which admits a two-dimensional homothety group.

\references 
\refjl{Barnes A and Rowlingson R R  1990}{\CQG}{7}{1721--31}
\refjl{Barnes A 1998}{\CQG}{15}{3061--70}
\refjl{Bradley M  1986}{\CQG}{3}{317--34}
\refjl{Edgar S B and Ludwig G  1997}{\CQG}{14}{L65--8}
\refjl{Edgar S B and Vickers J A  1999}{\CQG}{16}{589--604}
\refbk{Ehlers J and Kundt W 1962}{Gravitation: an introduction to
current research,  ed. L Witten}{Wiley, New York}
\refbk{Hearn A C 1995}{Reduce User's Manual, Version 3.6}{Rand, Santa
Monica, CA} 
\refjl{Karlhede A 1980}{\GRG}{12}{693--707}
\refjl{Karlhede A and MacCallum M A H 1982}{\GRG}{14}{673--82}
\refjl{Koutras A 1992}{\CQG}{9}{L143--5}
\refjl{Ludwig G and Edgar S B 2000}{\CQG}{17}{1683--1705}
\refjl{Pollney D, Skea J E F and d'Inverno R A 2000}{\CQG}{17}{2885--2902}
\refjl{Seixas W 1992}{\CQG}{9}{225--38}
\refjl{Skea  J E F  1997}{\CQG}{14}{2393-2404}
\refjl{Stephani H 1967}{Commun. Math. Phys.}{4}{137--42}
\refjl{Wils P  1989}{\CQG}{6}{1243--51}
\refbk{{\AA}man J E 1987}{Manual for CLASSI: classification program
for geometries in general relativity}{University of Stockholm, Institute of
Theoretical Physics  technical report}
\bye